\documentclass[twoside]{mhd}
\usepackage{graphicx}
\usepackage{bm}
\mhdhead{40}{1}{1}	

\title{DRESDYN - A new facility for MHD experiments\\ with liquid sodium}

\author{F. Stefani, S. Eckert, G. Gerbeth, A. Giesecke, Th. Gundrum,\\
C. Steglich, T. Weier, B. Wustmann}

\institute{Helmholtz-Zentrum Dresden-Rossendorf, 
P.O. Box 510119, D-01314 Dresden, Germany } 

\begin{document}
\maketitle


\begin{abstract}
The DREsden Sodium facility for DYNamo and thermohydraulic 
studies (DRESDYN) is intended as a platform 
both for large scale experiments related to 
geo- and astrophysics as well as  for experiments 
related to thermohydraulic and safety aspects of 
liquid metal batteries and 
liquid metal fast reactors. The most ambitious 
projects in the framework of DRESDYN are 
a homogeneous hydromagnetic dynamo driven solely by 
precession and a large Taylor-Couette type experiment for the combined 
investigation of the magnetorotational instability and the Tayler 
instability. 
In this paper we give a short summary about the 
ongoing preparations and 
delineate the next steps for the realization of DRESDYN. 
\end{abstract}


\section*{Introduction}
Beginning with the early considerations of Bevir \cite{BEVIR}, 
Pierson \cite{PIERSON}, and Steenbeck \cite{STEENBECK}, there has 
always been a tight connection between experimental dynamo studies and 
research related to liquid metal fast reactors (LMFR), in particular
to sodium fast rectors (SFR). For example, a precursor 
of the Riga dynamo experiment \cite{RIGA1} had been carried out in 1987 at a 
test facility for SFR pumps in Leningrad \cite{LENINGRAD}. Later, 
Alemany et al. \cite{ALEMANY} studied the possibility of self-excitation in the 
pumps and cores of the French Ph\'{e}nix and Superph\'{e}nix reactors. With 
the DRESDYN project we intend to resume this tradition by 
setting-up a new research infrastructure for liquid sodium experiments 
related both to the origin and action of cosmic magnetic fields (with
possible applications for the construction of large-scale 
liquid metal batteries) as 
well as to safety aspects of LMFR's.
First, we present the plans for a precession driven dynamo experiment and
a Taylor-Couette type experiment for the combined investigation
of the magnetorotational instability (MRI) and the Tayler
instability (TI). The destructive effect of the TI on the layered 
stratification in envisioned large-scale liquid metal batteries, 
and a possible provision to avoid
this effect, will then be discussed. The paper closes with  delineating 
further experiments related to In-Service-Inspection problems 
for LMFR's.

\section{Experiments with geo- and astrophysical background}

\subsection{Precession driven dynamo}
The most ambitious project within the framework of 
DRESDYN is a large scale precession dynamo experiment. 
Precession has been discussed since 
long as an, at least complementary, 
energy source of the geodynamo 
\cite{MALKUS,GANS,VANYO,TILGNER,SHALIMOV,JACQUESPRE}. 
This idea is supported by paleomagnetic measurements
that have revealed a modulation of the geomagnetic field intensity 
by the 100 kyr Milankovi\'{c} cycle of the 
Earth's orbit eccentricity and by the corresponding 41 kyr 
cycle of the Earth's axis obliquity \cite{CHANNELL}.
The 100 kyr cycle is also known to influence the
reversal statistics 
of the geomagnetic field, an effect that is very probably 
being enhanced by stochastic resonance 
\cite{MICHELIS,STOCHASTIC}. 
In this context there is an interesting 
correlation of geomagnetic field variations with climate changes,
clearly demonstrated for the last 5000 years \cite{KNUDSEN}, 
and perhaps also existing for the sequence of ice ages 
\cite{FRANZOSEN}, although the causal mechanism of how exactly 
the Milankovi\'{c} cycles affect both the climate and 
the magnetic field is far from being settled. 
In an early paper, 
Doake had speculated that the increasing ice sheets would change 
the moment of the inertia of the
Earth, thereby influencing the geodynamo by a modified rotation 
period \cite{DOAKE}.  
However it could be worthwhile to
investigate also the reverse causal chain that changing Earth's orbit 
parameters lead, in the first instance, to a modified geodynamo field 
which, in turn, could then influence
the climate by modifying the shield against cosmic ray flux
(which was recently discussed as a key climate driver through 
cloud formation \cite{SCHERER,SVENSMARK}).

\begin{figure}[t]
\begin{center}
\unitlength=\textwidth
\includegraphics[width=0.85\textwidth]{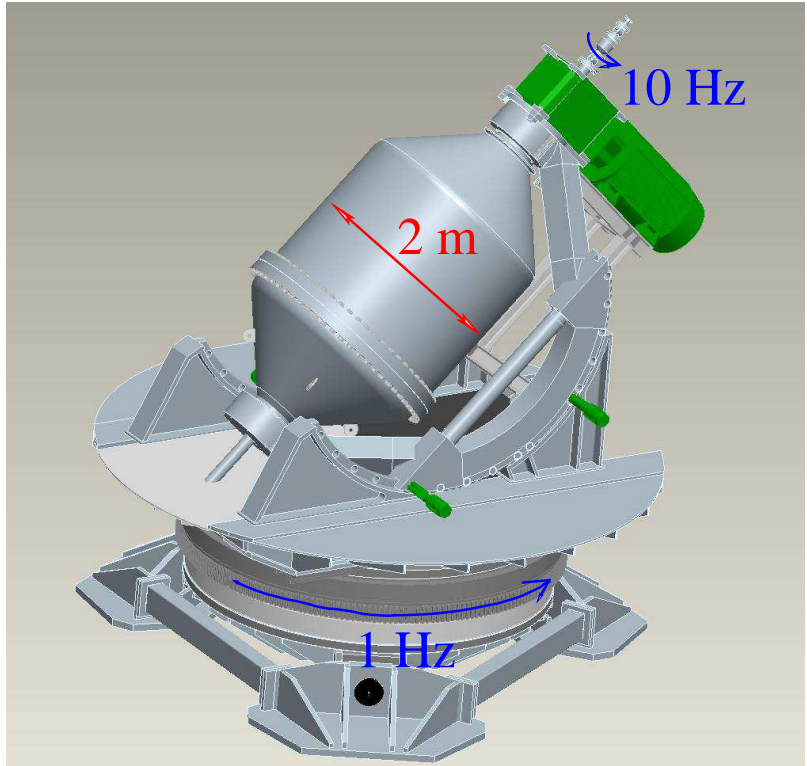}
\caption{Sketch of the planned large scale precession driven 
dynamo experiment with the envisioned sizes and rotation rates.}
\end{center}
\end{figure}   
 
Apart from this general geophysical motivation, a precession driven 
dynamo experiment is also interesting from the narrower 
magnetohydrodynamic point of view. 
In comparison with the previous experiments 
in Riga \cite{RIGA1}, Karlsruhe 
\cite{KARLSRUHE} and Cadarache
\cite{CADARACHE}, a precession experiment
would represent a {\it homogeneous} dynamo par excellence.
Containing only a homogeneous fluid rotating around two axes, 
it would neither contain any propeller, as in Riga, nor any 
assembly of 
guiding tubes, as in Karlsruhe, nor any soft-iron material 
(which is crucial for the low critical magnetic Reynolds number
and the  close to axisymmetric 
eigenmode in the Cadarache experiment
\cite{GIESECKE1}).

\begin{figure}[t]
\begin{center}
\unitlength=\textwidth
\includegraphics[width=0.75\textwidth]{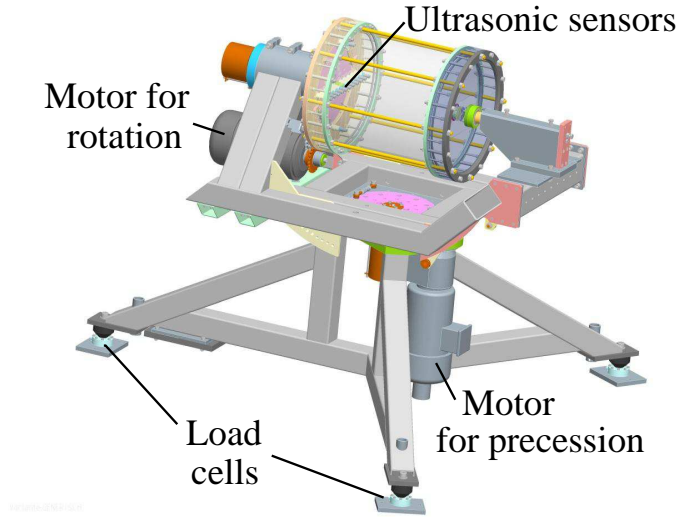}
\caption{Drawing of the 1:6 scale water precession experiment for the determination
of velocity field, the motor powers, and the torques on the basement
for various driving conditions.}
\end{center}
\end{figure}

The central part of the envisioned precession dynamo experiment
(see Fig. 1 for a preliminary draft) will be a cylindrical 
vessel of approximately 2 m diameter and length, 
rotating with up to 10 Hz 
around its symmetry axis, and with up to 1 Hz around another axis
whose angle to the first axis can be varied between 
90$^{\circ}$ and 45$^{\circ}$. 
The inner cylindrical shell (made of copper) is immersed 
into a larger 
cylindrical stainless container with conical end parts
that can later also be used to house alternative inner shells,
e.g. in the form of ellipsoids.

The mechanical and safety demands for such a large 
scale sodium experiment are tremendous. With a total mass of 
approximately 20 tons (including the sodium in the 
conical end parts and the stainless steel 
walls) the sodium filled cylinder will have a moment of inertia of around 
10$^4$ kg m$^2$. With a rotation rate of 10 Hz and a precession rate of 
1 Hz, this 
amounts to a gyroscopic moment of 5x10$^6$ Nm which 
requires an extremely massive basement and a careful
avoidance of resonances. Further, the experiment must be fenced 
by a containment which can be quickly flooded with Argon in case 
of a sodium accident.

In order to determine various crucial parameters for the design of this 
large-scale sodium experiment, we have started a series of experiments at
a smaller (scale 1:6) water precession experiment which is
shown in Fig. 2. This small water experiment is 
similar to the ATER experiment guided by J. L\'{e}orat \cite{JACQUES}, but 
comprises some special measuring devices. Most important 
for the later sodium experiment 
is the determination of the torques and motor powers needed to drive the 
rotation of the cylinder and the turntable, and of the gyroscopic torques 
acting on the basement. Those measurements have already confirmed the 
expected sharp transition between a  laminar flow and a 
turbulent flow which in our case occurs at a precession ratio of 
around 0.07. At this point, the needed motor power increases sharply, 
as indicated in Fig. 3a.

Concerning the flow field determination, up to present we 
have only installed a number of ultrasonic sensors for the determination of the 
axial velocity component. 
For a (rather low) rotation rate of 0.2 Hz and a precession rate of 
0.01 Hz, Fig. 3b shows first 
results of the axial velocity measured by Ultrasonic 
Doppler Velocimetry (UDV). 

\begin{figure}[t]
\begin{center}
\unitlength=\textwidth
\includegraphics[width=0.99\textwidth]{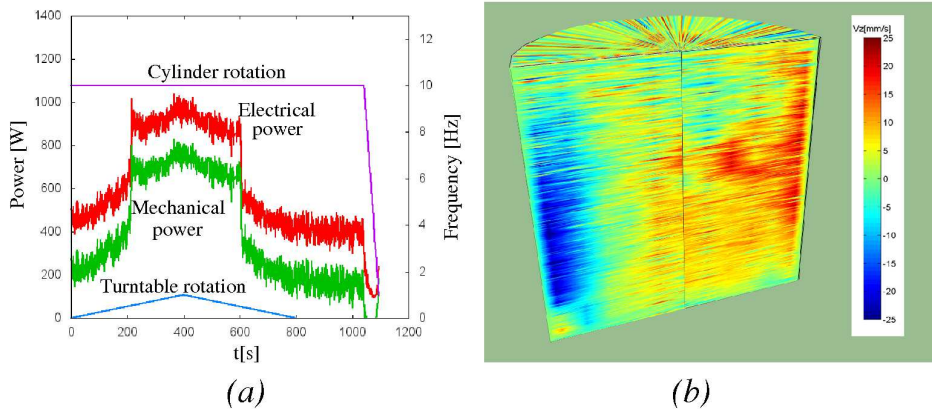}
\caption{First results obtained with the water precession experiment. (a)
Motor power (electrical and mechanical) in dependence on the
precession rate. Note the sudden jump of the power for a precession ratio 
of approximately 0.07 which indicates the transition between the laminar and
the turbulent flow regime. (b) Axial velocity component for half a rotation, measured 
by 6 Ultrasonic sensors at a 
rotation rate of 0.2 Hz and a precession rate 0.01 Hz. 
The sign change indicates the typical Kelvin mode (m=1) structure.}
\end{center}
\end{figure}

\subsection{MRI/TI experiment}

The second liquid sodium experiment with geo- and astrophysical 
motivation will be a large-scale  Taylor-Couette-Experiment (Fig. 4)
with a diameter of approximately 1 m and a height of 2.5 m, 
wrapped by a coil that produces a strong vertical magnetic field, 
and supplemented by technical means to guide independent 
electrical currents through a bore in the center 
and through the liquid sodium between the inner and outer cylinder. 

In the following this experiment will be called the
''MRI/TI experiment'', indicating 
its astrophysical background in terms of the Magnetorotational 
instability (MRI) and the Tayler instability (TI).
The MRI had been discussed, for the first time, by 
Velikhov in 1959 \cite{VELIKHOV},
and was then rediscovered and applied to accretion 
disk physics by Balbus and 
Hawley in 1991 \cite{BAHA}. Nowadays, MRI 
is thought to play a crucial role in 
enabling angular momentum transport by destabilizing the
Keplerian rotation profiles of
accretion disks that would otherwise be hydrodynamically 
stable.

With this large-scale MRI/TI experiment we plan to extend our 
previous investigations \cite{MRI1,MRI2} 
on the helical version of MRI (HMRI) to the realm of the 
standard MRI (SMRI). A particular focus will be on the 
interesting mode transitions between 
the HMRI and the SMRI, as  discussed recently in 
\cite{KIRILLOV1,KIRILLOV2}.

In addition to this, the experiment will also allow to study the kink-type 
Tayler instability (TI) \cite{TAYLER,SPRUIT} and its transitions to the
azimuthal MRI (AMRI) \cite{AMRI}.
Recent experimental work on TI with the eutectic alloy GaInSn has 
revealed a complicated interplay of the very TI with large scale 
convection  effects due to Joule heating \cite{SEILMAYER}. With the envisioned sodium
experiment we intend to suppress those convection effects significantly.
On this basis, we see a much better chance 
to study the numerically predicted intrinsic saturation effects of 
TI by means of an increased turbulent resistivity \cite{GELLERT}.

\begin{figure}[t]
\begin{center}
\unitlength=\textwidth
\includegraphics[width=0.95\textwidth]{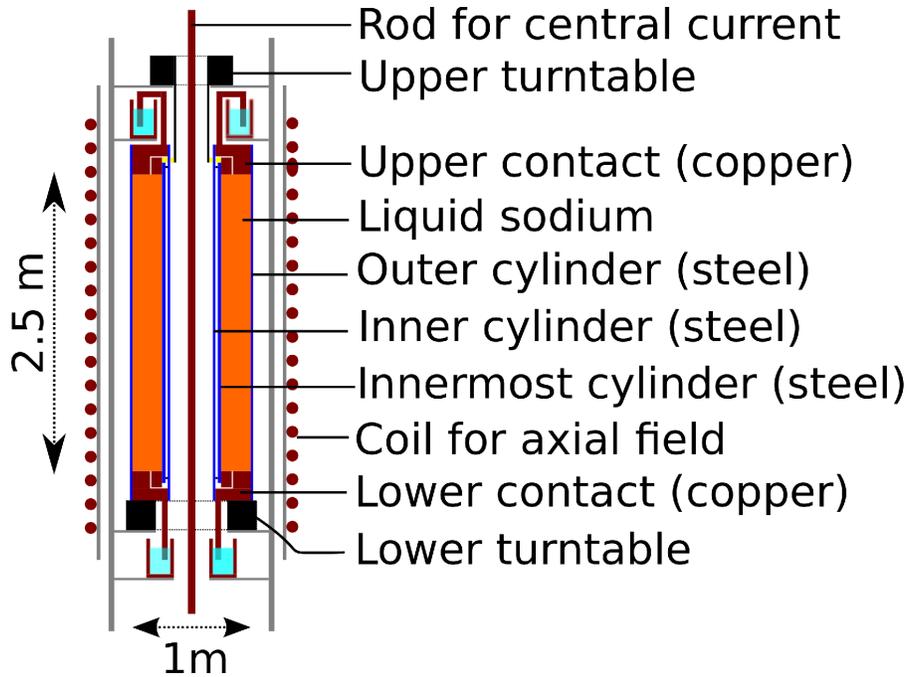}
\caption{Sketch of the Taylor-Couette experiment for the combined study of
MRI and TI.}
\end{center}
\end{figure}

\section{TI and liquid metal batteries}

Tightly connected with the research on TI, we will also investigate
liquid metal batteries that have recently 
been proposed as cheap devices for the storage of 
the highly fluctuating renewable energies. 

Typically, such a battery would consist of a 
self-assembling stratification of a 
heavy liquid half-metal (e.g. Bi, Sb) at the bottom, an 
appropriate molten salt as electrolyte in the middle, 
and a light alkaline or earth alkaline metal (e.g. Na, Mg) at the top. 
The functioning of a small version of this type of battery has already been 
verified \cite{BRADWELL}. However,  
for large-scale batteries (which are the only interesting ones 
in terms of economic competitiveness) 
the occurrence of TI can be easily imagined to represent a 
serious problem for the integrity of the stratification
(see Fig. 5a).

In our experiments, we will focus on various ways to avoid TI in such
configurations, e.g. by using a  return current through the center of the 
battery (see Fig. 5b), as it was proposed in a recent paper \cite{ECM}.

\begin{figure}[t]
\begin{center}
\unitlength=\textwidth
\includegraphics[width=0.75\textwidth]{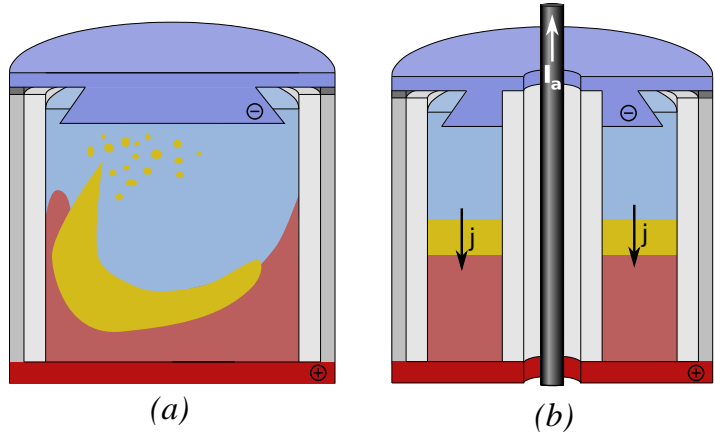}
\caption{Liquid metal batteries. (a) Illustration of the expected action of TI 
on the self-assembled stratification in a liquid metal battery. (b)
A simple provision to avoid TI by returning the charging/discharging 
current through a bore in the center.}
\end{center}
\end{figure}

\section{Experiments related to liquid metal fast reactors (LMFR)}

The safe and reliable operation of liquid metal systems in 
innovative reactor concepts like sodium fast reactors (SFR)
or transmutation systems based on lead-bismuth cooled reactors 
(LBFR) requires appropriate measuring systems and control units, both for the 
liquid metal single-phase flow as well as for gas bubble liquid 
metal two-phase flows. Hence, there is a growing need for 
small and medium sized liquid sodium experiments to study various 
thermo-hydraulic and safety aspects of SFR's, comprising sodium boiling, 
argon entrainment, bubble detection, sodium flow metering and many more 
\cite{TENCHINE}.

\begin{figure}[t]
\begin{center}
\unitlength=\textwidth
\includegraphics[width=0.75\textwidth]{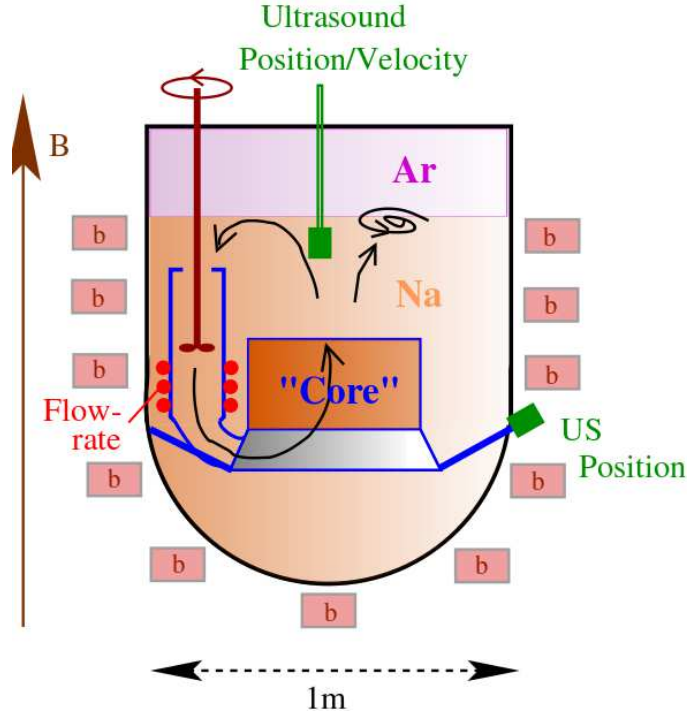}
\caption{Sketch of the planned In-Sevice-Inspection experiment. The 
magnetic field ''B'' is the measuring field needed for the application
of the Contactless Inductive Flow Tomography (CIFT), and the small boxes 
with ''b'' indicate the magnetic field sensors for measuring the flow 
induced magnetic fields. }
\end{center}
\end{figure}

A significant portion of the LMFR-related experiments in the framework of 
DRESDYN will be conducted at an In-Service-Inspection (ISI) facility the 
principle sketch of which is shown in Fig 6. Basically, it will consist of a 
heated stainless steel vessel with a diameter of approximately 1 m that is 
filled with liquid sodium covered by argon. The internal components will comprise 
a simple mock-up of a reactor core and a primary pump. The main goal of this 
facility is to test a variety of measurement techniques for the position of 
internal components, for flow velocities and argon bubble detection. The 
latter will also include experiments on Argon entrainment on the free surface.
Particular attention will be paid to the application of the 
Contactless Inductive Flow Tomography (CIFT) for the flow
reconstruction in LMFR's. This technique has been
developed at HZDR during the last decade \cite{CIFT1,CIFT2}, and 
it was recently deployed for visualizing the flow structure 
in a physical model of the continuous casting of steel \cite{CIFT3}.
CIFT could be of particular value for Lead-Bismuth cooled 
transmutation systems, such as MYRRHA \cite{MYRRHA}. 
The lower ''cold pool'' of MYRRHA is characterized by a rather 
free velocity field which is very likely prone to flow
instabilities that could easily be detected by CIFT.

DRESDYN will also comprise a liquid sodium loop which 
will replace the presently existing one. This loop will 
contain various 
test sections, among them one section for the study of smart heat 
exchangers with intermediate heat transfer media.
The deployment of such smart heat exchangers in future SFR's could 
reduce significantly the risk of energetic sodium-water reactions.
Another suite of experiments will be devoted to the important problem of 
sodium boiling which is a key safety issue for SFR's due to their
positive reactivity coefficient. Starting with very small experiments 
at a flat wall, we 
plan to go over later to boiling experiments at rods or rod bundles. For the 
visualization of the boiling we plan to use the Mutual Inductance 
Tomography \cite{TERZIJA}, as well as X-ray radiography \cite{BODEN} and, 
in collaboration 
with another group at HZDR, the ultrafast X-ray tomography \cite{HAMPEL}.

Finally, with the advent of Oxide Dispersion Strengthened (ODS) steels 
as new promising nuclear reactor materials (because of their superior 
swelling resistance and excellent high temperature strength) \cite{KAITO},
we come back to our starting point of considering the 
possibility of magnetic-field self-excitations in SFR's. 
It has been shown that the use of ferritic or 
martensitic steels in the
core of large SFR's could indeed foster
magnetic field self-excitation \cite{PLUNIAN,SOTO}.
This is consistent with the key role of 
high magnetic permeability material for the
dynamo process in the Cadarache experiment \cite{GIESECKE1}.
Hence, it is certainly necessary to 
reconsider this point before new steel sorts can be 
utilized in large scale SFR's. Related experiments
on this topic are also envisaged in the framework of DRESDYN.

\section{Conclusions}

In this paper, we have discussed the motivation behind, and the
concrete plans for a number of experiments to be set-up in the 
framework of DRESDYN. The new building and the essential parts of the
experiments are expected to be ready in 2015. 
Apart from hosting the discussed experiments, 
DRESDYN is also meant as a general platform for further large-scale 
experiments, basically but not exclusively with liquid sodium. 
Proposals for  such experiments are, therefore, highly welcome.

\section*{Acknowledgments}
This research was supported by Deutsche Forschungsgemeinschaft 
(DFG) under grant STE 991/1-1 and in frame of the SFB 609 
"Electromagnetic Flow Control in Metallurgy, Crystal Growth and 
Electrochemistry". We thank Jacques L\'eorat for his proposal and his 
insistence to build a precession driven dynamo, and Canlong
Ma for assistence in the processing of data from the water precession
experiment.



\newcommand{\noopsort}[1]{}

\lastpageno

\end{document}